\documentclass[sigconf]{acmart}
\usepackage{graphicx}
\usepackage{subcaption}
\usepackage{booktabs} 
\usepackage[listings]{tcolorbox}
\usepackage{xcolor}
\usepackage{multirow}
\usepackage{hyperref}
\usepackage{url}
\usepackage{comment}
\usepackage{todonotes}
\usepackage{color}
\usepackage{balance}
\usepackage{tabularx}
\newcolumntype{b}{X}
\newcolumntype{s}{>{\hsize=.45\hsize}X}
\newcolumntype{x}{>{\hsize=.25\hsize}X}
\usepackage{algorithm}
\usepackage[noend]{algpseudocode}

\copyrightyear{2024} 
\acmYear{2024} 
\setcopyright{acmcopyright}
\acmPrice{0.00}
\acmDOI{10.1145/xxxxxxx.xxxxxxx}
\acmISBN{xxx-xxx-x-xx-xx-xx}

\begin{document}

\title[Ensemble Machine Learning Algorithms for Automatic Test Case Generation]{Introducing Ensemble Machine Learning Algorithms for Automatic Test Case Generation using Learning Based Testing}

\author{Sheikh Md. Mushfiqur Rahman}
    \affiliation{%
        \institution{Boise State University}
        \city{Boise}
        \state{ID}
        \country{USA}
}
\email{sheikhmdmushfiqu@u.boisestate.edu}

\author{Nasir U. Eisty}
\affiliation{%
	   \institution{Boise State University}
	   \city{Boise}
	   \state{ID}
	   \country{USA}
}
\email{nasireisty@boisestate.edu}


\begin{abstract}
 \textit{\textbf{Context}}: 
Ensemble methods are powerful machine learning algorithms that combine multiple models to enhance prediction capabilities and reduce generalization errors.
However, their potential to generate effective test cases for fault detection in a System Under Test (SUT) has not been extensively explored.
 \textit{\textbf{Objective}}: 
This study aims to systematically investigate the combination of ensemble methods and base classifiers for model inference in a Learning Based Testing (LBT) algorithm to generate fault-detecting test cases for SUTs as a proof of concept.
 \textit{\textbf{Method}}: 
We conduct a series of experiments on functions, generating effective test cases using different ensemble methods and classifier combinations for model inference in our proposed LBT method.
We then compare the test suites based on their mutation score.
 \textit{\textbf{Results}}: 
The results indicate that Boosting ensemble methods show overall better performance in generating effective test cases, and the proposed method is performing better than random generation. This analysis helps determine the appropriate ensemble methods for various types of functions.
 \textit{\textbf{Conclusions}}: 
By incorporating ensemble methods into the LBT, this research contributes to the understanding of how to leverage ensemble methods for effective test case generation.

\end{abstract}

%
%


\settopmatter{printfolios=false}

\maketitle

\section{Introduction}
\label{sec:Introduction}
Automated software testing with Machine Learning (ML) has garnered significant attention, with an initial focus of research being centered around Test Oracle generation~\cite{aggarwal2004neural,jin2008artificial,sangwan2011radial}, which involves creating reliable and accurate expected outputs or labels for a given set of test cases.
Furthermore, ML has significantly improved the automation of testing, enabling tasks that were previously considered nearly impossible to automate, such as testing android application gui~\cite{collins2021deep}, web pages~\cite{kamal2019enhancing} using deep learning models, highlighting the potential of ML in making testing processes more efficient.


In Black-box testing, selecting suitable test cases to detect faults is challenging because it's done without access to the program's source code. The process is often random, and covering the vast input space comprehensively can be difficult~\cite{zhauniarovich2015towards}.
Researchers have explored Reinforcement Learning and conventional supervised learning techniques to improve coverage and fault detection capabilities in generating test cases for Black-box systems, showing promise in addressing this challenge~\cite{esnaashari2021automation, papadopoulos2015black}. 
ML approaches enhance Black-box testing, improving its efficacy for more effective software testing without requiring access to the source code.

Learning Based Testing (LBT) offers a viable solution by integrating testing and model inference, inferring multiple SUT models with a small set of test cases iteratively. Test cases challenging inferred model predictions are added to the suite and used for model retraining~\cite{walkinshaw2010increasing}. The technique focuses on selecting test cases likely to detect faults by emphasizing input/output relations contradicting the inferred model~\cite{papadopoulos2015black, walkinshaw2017uncertainty}. 
LBT's incremental learning of the SUT sets it apart, enabling a scalable and efficient process~\cite{feng2013case}. 
Moreover, Meinke and Niu~\cite{meinke2010learning} demonstrated that LBT can significantly outperform random testing regarding speed when detecting errors in a SUT.
Finally, this method also seems to solve the issue of testing insufficiency that arises from the source code drive testing methods, which commonly have syntax-centric views~\cite{fraser2015assessing}.

Various techniques, such as genetic algorithms and state machines, have been applied in LBT for model inference~\cite{esnaashari2021automation, raffelt2005learnlib}. 
However, to introduce a novel approach, this paper introduces a novel approach by implementing a different combination of ensemble methods and base classifiers within the LBT algorithm for model inference, using a specification-based test case generation technique. Ensemble methods aggregate predictions from multiple classifiers using weighted voting to classify new data points~\cite{dietterich2000ensemble}. We utilize ensemble classifiers' diversity and Z3's constraint-solving capability to generate and select effective test cases. As far as we know, no LBT method has integrated ensemble methods into the model inference technique.


To guide this study, we formulate the following research questions that group around the core aspects of our investigation:

\textbf{RQ1: Which combination of ensembles performs better for a classification function?}

\textbf{RQ2: Which combination of ensembles performs better for a numeric function?}

\textbf{RQ3: Which combination of ensembles performs better overall?}

The data and artifacts of this paper can be accessed \textcolor{blue}{\href{https://figshare.com/s/9ef8ee16f157864b4802}{here}}\footnote{https://figshare.com/s/9ef8ee16f157864b4802}.


\begin{figure}[htbp]
\centering
\subfloat[ELBT algorithm]{\includegraphics[width=\columnwidth]{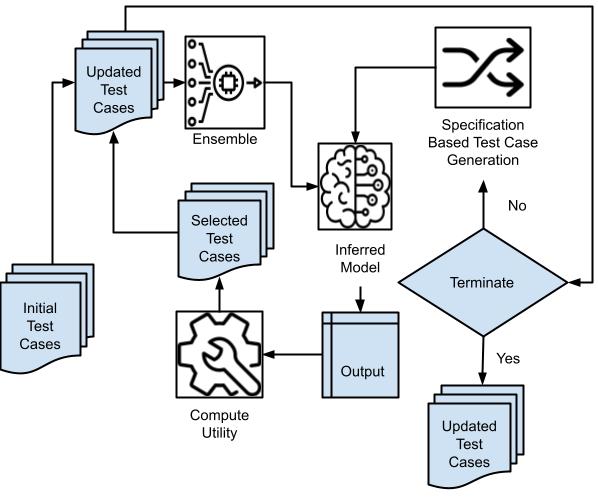}}
\
\subfloat[Mutation Testing]{\includegraphics[width=\columnwidth,height=2.5cm]{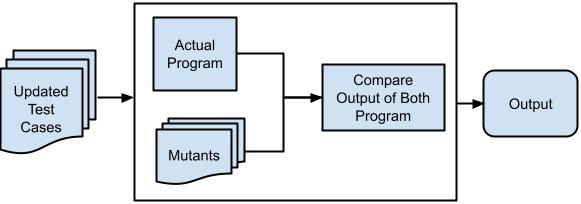}}
\caption{Process Diagram}
\label{fig:process_diagram}
\end{figure}

\section{Algorithms \& Evaluation}
\label{sec:Algorithms}

\subsection{\textbf{Learning Based Testing (LBT)}}
\label{tbc_algorithm_description}


Weyuker first introduced the concept of LBT ~\cite{weyuker1983assessing}, where testing is portrayed as an inference process wherein testers try to discern software attributes by examining its response to distinct inputs. The LBT approach uses a concise test set for a SUT, labeled as $P$. It repeatedly produces a program, $p'$, that is part of $P$, conforming with the test set. This approach looks for a unique input differentiating $P$ from $p'$, and this continues until only $P$ can be derived from the generated examples~\cite{bergadano1996testing}. 
The primary objective of LBT is to enhance specification-based black box testing~\cite{meinke2011learning}.
Feng et al.~\cite{feng2013case}  demonstrated that utilizing LBT enables generating a considerable number of effective test cases by integrating the model-checking algorithm with an iterative model inference technique. Budd and Angluin~\cite{budd1982two} first highlighted two fundamental issues in program testing, and here is how LBT resolves them: 

\begin{itemize}
    \item Selecting effective test cases: LBT selects test cases efficiently by approximating the SUT and identifying relevant cases likely to uncover defects. 
    \item Determining a stopping criterion for adequate testing: LBT establishes a stopping criterion to determine when testing is adequate, optimizing the process to achieve sufficient test coverage without unnecessary resource expenditure.
\end{itemize}

For this experiment, we implement the proposed ELBT algorithm, an LBT algorithm for automatic test generation, shown in Figure~\ref{fig:process_diagram}(a). 
Our implementation uses various ensemble methods and base classifiers for model inference. Additionally, we introduce a specification-based test case generation technique using an SMT solver, which solves decision problems to determine the satisfiability of logical formulas with respect to specific mathematical theories~\cite{de2008z3}. 
To implement the specification-based test case generation technique, we convert the program specifications and known conditions of the SUTs into formulas that are supported by the SMT solver, and using those, the SMT solver will generate inputs for selection. 
To understand test case generation using Ensemble models as the model inference technique 
, we explain the steps:

\begin{enumerate}
    \item Initiate the process with an initial small set of test cases generated by the proposed specification-based test case generator and train the ensemble models using this test set.
    
    \item For a fixed number of iterations:
    \begin{itemize}
        \item Generate test inputs from the proposed test generator.
        \item Calculate the Utility of all test inputs in the generated test set using the inferred model.
        \item Select the test input with the highest utility.
        \item Execute the chosen test input using the SUT and incorporate it into the existing set of test cases.
        \item Train the models using the updated test set.
    \end{itemize}
    \item Return to step 2 and repeat the entire process.
    \item Continue running the full process until the desired number of test cases is selected.

\end{enumerate}



\subsection{\textbf{Ensemble Method}}


An ML ensemble combines individual models' outputs through weighted or unweighted voting~\cite{dietterich2000ensemble}. The two primary forms of ensemble methods used are: 1) Bagging
~\cite{breiman1996bagging}, a popular ensemble algorithm which helps to reduce the variance of the individual models by introducing randomness through bootstrapping and
2) Boosting~\cite{schapire1990strength}, which 
was introduced to improve the performance of a weak ML algorithm by reducing bias.
Ensemble methods like Bagging and Boosting improve ML models but differ in their training processes. Bagging trains models in parallel with equal sample weights, while Boosting uses a sequential approach, assigning higher weights to misclassified samples to focus on difficult instances. Both methods are versatile and applicable to various learning algorithms. Ensembles excel over single models, especially with complex or noisy data~\cite{fernandez2014we},
enhancing ML system performance.

\subsection{\textbf{System Under Test (SUT)}}


Our experiment is conducted to generate test cases for two distinct SUTs: 
1) A classification function, the classic Triangle Classification function, determines whether three input parameters represent an equilateral, isosceles, or invalid triangle. 
Notably, various testing techniques such as search-based, white box, and black box testing have previously utilized this unction as the SUT for test case generation~\cite{mcminn2004search, esnaashari2021automation, fraser2015assessing}. 
2) A numeric function called the Find Middle function, which takes three numbers as input and returns the middle number. This function has been a demo program in several test case and model generation techniques~\cite{fraser2015assessing, ghani2008strengthening}. 



\subsection{\textbf{Mutation Testing}}

Mutation testing evaluates the quality of a test suite by introducing slight alterations (mutations) to the program's code and checking if the tests can detect these changes. Various mutation frameworks are available for program testing, including PIT~\cite{coles2016pit}, DeepMutation~\cite{tufano2020deepmutation}, IBIR~\cite{khanfir2023ibir}, and $\mu$Bert~\cite{khanfir2023efficient}. ML algorithms have also effectively utilized mutation testing for bug detection~\cite{cheng2018manifesting}. 
In metamorphic relation detection, mutation testing has been actively applied, showcasing its versatility and utility across different domains~\cite{nair2019leveraging}. We selected mutation testing as the evaluation metric for the generated test suite due to its ability to assess effectiveness with reduced computational cost~\cite{jia2010analysis}. Additionally, many LBT-based black box testing methods have successfully employed mutation testing to measure the quality of their generated test suites~\cite{walkinshaw2017uncertainty,papadopoulos2015black}.

\section{Research Methodology}
\label{sec:Methodology}
This section presents an overview of our research methodology, encompassing test case selection and mutation testing. 
The complete process is visualized in Figure~\ref{fig:process_diagram}. 

\subsection{\textbf{Automatic Test Case Generation}}
In this experiment, we focus on 
a small set of widely used ensemble methods that use the core concept of bagging and boosting method and are available in scikit-learn library for implementation. 
In conjunction with the ensemble methods, we employ a set of conventional supervised ML algorithms as the base learners or estimators. 
For clarity, we use the terms ``estimator" and ``learner" interchangeably. 
Each weak classifier of the ensemble is of the same type as the base estimator. 
We create an ensemble combination by configuring the base estimators for a particular ensemble method.

\subsubsection{\textbf{Implementation}}
To create the dataset of test cases, we generate inputs by using the Z3, an SMT solver specifically designed to address challenges encountered in software verification and software analysis~\cite{de2008z3}. Finally, the generated inputs are labeled using the previously discussed SUTs for this experiment. To create test case inputs, all the known specifications of the SUTs are converted into formulas, which are supported by the Z3 solver. For example, a known specification of one of the SUTs, the triangle classification function, is that if all the sides are equal, then the triangle is an Equilateral triangle. For this specification, apart from the valid triangle constraints, we added a constraint is the Z3 solver as ($x$==$y$,$y$==$z$,$z$==$z$) where $x$,$y$,$z$ are 3 sides of a triangle.

As outlined in Figure~\ref{fig:process_diagram}, we select test cases with the highest utility from the generated input set. 
Figure~\ref{fig:process_diagram} illustrates that the ensemble's predictions for the inputs generated by the proposed input generator are used to calculate the utility of these test cases. 
In an ensemble, the outputs of the models are valuable if there is disagreement in their predictions for an input~\cite{krogh1994neural}. 
We choose the input among the generated inputs with the highest disagreement in their outputs as the test case to be collected for model training and testing the SUT in our algorithm. 
In summary, this algorithm selects test cases with the highest diversity in ensemble predictions. In another way, using the diversity, we are prioritizing which specification needs to be tested in black box testing. In this way, the cost of testing can be reduced by testing only a set of specifications of the SUT rather than testing the full SUT. In addition, The random test suite, against which our test suite generated by ELBT will be compared, is generated using Python's built-in random module. 

\subsubsection{\textbf{Computing Test Case Utility:}}
The LBT method operates by simultaneously improving the SUT approximation through model inference and testing the SUT. Test cases are selected based on ensemble diversity, as diversity has been shown to be crucial for accurate ensembles~\cite{kuncheva2003measures}. 
The amount of disagreement among the ensemble members is referred to as the diversity of the ensemble, which is measured as the probability that a random ensemble prediction on a random example will disagree with the prediction of the complete ensemble~\cite{melville2004diverse}. Previous research has shown that there is a correlation between diversity and ensemble error reduction~\cite{melville2005creating}. 
In short, diversity in the ensemble leads to improved classification accuracy by creating uncorrelated classifications~\cite{hu2001using}. 
So, we use the diversity of the ensemble methods to compute the utility of the random test cases as depicted in figure \ref{fig:process_diagram}.

Each ensemble method employs distinct approaches to introduce diversity within its collection of models. Bagging generates diversity by training classifiers on different subsets of data, while Boosting adjusts data distributions for subsequent classifiers. Random Forests achieve diversity through training on diverse data subsets and feature sets~\cite{melville2005creating}.

Kuncheva and Whitaker~\cite{kuncheva2003measures} have explored several methods to assess the diversity of predictions by the ensemble. For numeric functions, the standard deviation can be used for this purpose. 
However, we use Mean Absolute Deviation (MAD) to measure the diversity of outputs for inputs, as it avoids the impact of potential ``data spikes" wherein one model produces an outlier compared to the others~\cite{walkinshaw2017uncertainty}. MAD quantifies how much individual data points deviate, on average, from the mean of the data set. 
The diversity of the ensemble of size $n$ for input $x$ in a numeric function can be described by Equation~\ref{eq:variance}:

\begin{equation}\label{eq:variance}
\text{d}(x) = \frac{1}{n} \sum_{i=1}^{n} (M_i(x) - \mu)
\end{equation}

Here, $M_i(x)$ is the output from the $i^{th}$ classifier and $\mu$ is the mean value of the predictions of the ensemble of size $n$.

In our classification functions, we employ a formula inspired by two equations used by Melville and Mooney~\cite{melville2005creating, melville2003constructing} to calculate the ensemble's diversity for a test suite. 
According to their equations, the diversity of the $i^{th}$ classifier for input $x$ is defined as follows:

\begin{equation}\label{eq:di}
d_i(x) = \begin{cases}
0, & \text{if } M_i(x) = M^*(x) \\
1, & \text{otherwise}
\end{cases}
\end{equation}

where \(M_i(x)\) is the prediction of the $i^{th}$ classifier and \(M^*(x)\) represents the prediction of the entire ensemble. This measure is logical as it quantifies the likelihood of a classifier within an ensemble disagreeing with the overall prediction made by the ensemble~\cite{melville2005creating}.

As we want to calculate diversity for an ensemble for a specific test case, not a test suite, for an input $x$, the diversity of the ensemble of size $n$ will be calculated as:
\begin{equation}
   d(x) = \frac{1}{n}\sum_{i=1}^n  d_i(x)
\end{equation}

In summary, our proposed LBT method calculates the diversity of each classifier in the ensemble to determine the overall ensemble diversity for an input. The input with the highest diversity is selected, but if all inputs have the same diversity, a random selection is made. This approach ensures effective test case selection even when diversity scores are equal, as random selection in one iteration influences diversity differences in subsequent iterations.

\subsection{\textbf{Experiment}}
To facilitate easier reference, we use shortened names for both ensemble methods and base estimators. For example, the combination of BaggingClassifier (BC) with DecisionTreeClassifier (DTC) is denoted as BC-DTC. However, when RandomForestClassifier (RFC) or ExtraTreesClassifier (ETC) is used as the ensemble method, DecisionTreeClassifier (DTC) is automatically set as the base estimator, and only the shortened names of the ensemble methods are used. For instance, if ExtraTreesClassifier is used as the ensemble method, its combination is expressed as ETC. This format simplifies comprehension moving forward. 
\subsubsection{\textbf{Ensemble for Triangle Classification Function}}

we conduct test case generation for the triangle classification function. 
The experiment is carried out for nine sets of ensemble combinations.

\textbf{\textit{Bagging}.}
As Bagging methods, we utilize three classifier ensemble methods that are available in the scikit-learn library.: BaggingClassifier (BC), RandomForestClassifier (RFC), and ExtraTreesClassifier (ETC). For BC, three classifiers are used as Base Estimators to create the models: DecisionTreeClassifier (DTC), LogisticRegression (LOR), and RFC. 
Among these, DTC and LOR are conventional supervised ML methods. 
However, RFC is a Bagging method that uses DTC as its default Base Estimator. 
Similarly, ETC is a Bagging method similar to RFC, utilizing DTC as its base estimator.


\textbf{\textit{Boosting}.}
For Boosting method, we implement AdaBoostClassifier (ABC) and GradientBoostingClassifier (GBC) as the main ensemble methods.
For GBC, the default base estimator is DecisionTreeClassifier (DTC). 
As for ABC, we utilize DecisionTreeClassifier (DTC) and LogisticRegression (LOR) from the scikit-learn library as the base estimators. 
Additionally, RandomForestClassifier (RFC) is also used as a base estimator for ABC. The combination ABC-RFC is unique as it combines a Bagging method as the base estimator for a Boosting method. 
This is the only combination where we utilized both Bagging and Boosting.


\subsubsection{\textbf{Ensemble for Find Middle Function}}
Similarly, we choose the find middle function as the numeric function and employ nine sets of regression ensemble combinations for model inference in the proposed LBT method.

\textbf{\textit{Bagging}.}
For Bagging methods, we utilize three ensemble methods: BaggingRegressor (BR), RandomForestRegressor (RFR) and ExtraTreesRegressor (ETR), all of which are available in the scikit-learn library. 
As the base estimator for BR, we employ DecisionTreeRegression (DTR),  LinearRegression (LIR), and RFR. 
Notably, the BR-RFR combination is unique as both the ensemble BR and base estimator RFR are bagging methods.
Additionally, both RFR and ETR have DTR as base estimator by default.

\textbf{\textit{Boosting}.}
In the Boosting methods, we deploy AdaBoostRegressor (ABR) and GradientBoostingRegressor (GBR) as the ensemble methods. 
For the ABR ensemble method, we use
DecisionTreeRegressor (DTR), LinearRegression (LIR), and RandomForestRegressor (RFR) as the base estimator. 
Notably, the ABR-RFR combination is unique, as it involves using a Bagging method as a base estimator for a Boosting ensemble method.

\subsection{\textbf{Evaluation}}

As depicted in Figure~\ref{fig:process_diagram}(b), we proceed with mutation testing after generating the targeted number of test cases to evaluate the effectiveness of the test cases. For this purpose, we use the $\mu$BERT framework~\cite{khanfir2023efficient}, which leverages CodeBERT~\cite{feng2020codebert}, a pre-trained language model, to create mutants by masking and replacing tokens in a program's code. 
We select $\mu$BERT as the mutation framework for this experiment because khanfir et al.~\cite{khanfir2023efficient} demonstrated that it excels in generating fault detection test suites with minimal effort, surpassing state-of-the-art techniques like PiTest~\cite{coles2016pit} in terms of performance.

To ensure unbiased testing results, we exclude mutants that trigger exceptions during the experiment. Using valid and executable test cases helps detect and eliminate these exception-triggering mutants effectively. Relying on easily killable mutants could inflate mutation scores for ineffective ensemble methods, leading to biased outcomes. After generating mutants of the SUTs using the $\mu$BERT framework, we apply various ensemble methods on the ELBT algorithm to generate test cases. We observe the mutation score per iteration, calculated as the percentage of mutants killed by the test suite, reflecting its quality.


\section{Results}
\label{sec:Results}
\begin{figure*}[htbp]
\centering
\subfloat[]{\includegraphics[width=0.33\textwidth, height=4.5cm]{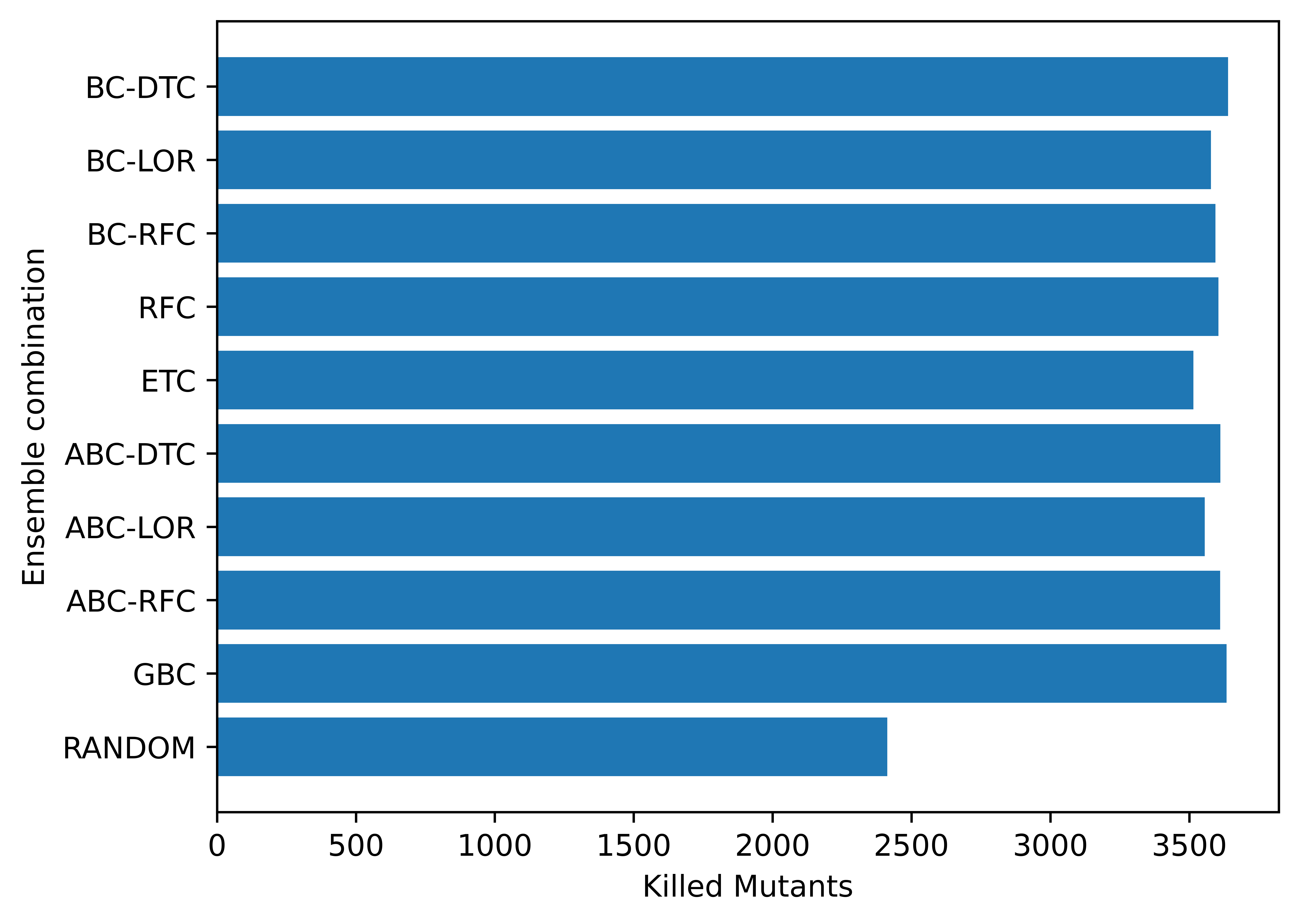}}
\subfloat[]{\includegraphics[width=0.33\textwidth]{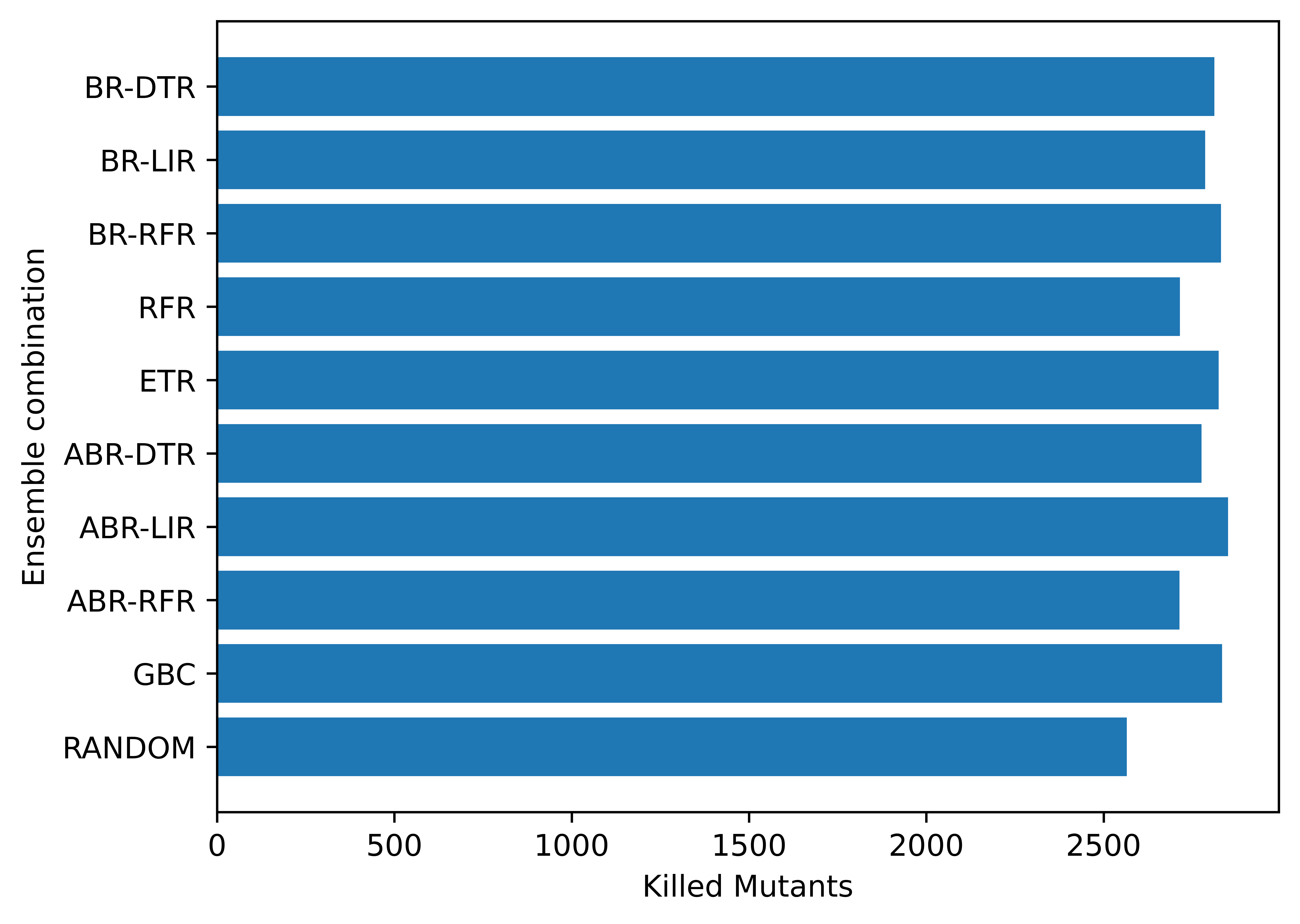}}
\
\subfloat[]{\includegraphics[width=0.33\textwidth,height=4.5cm]{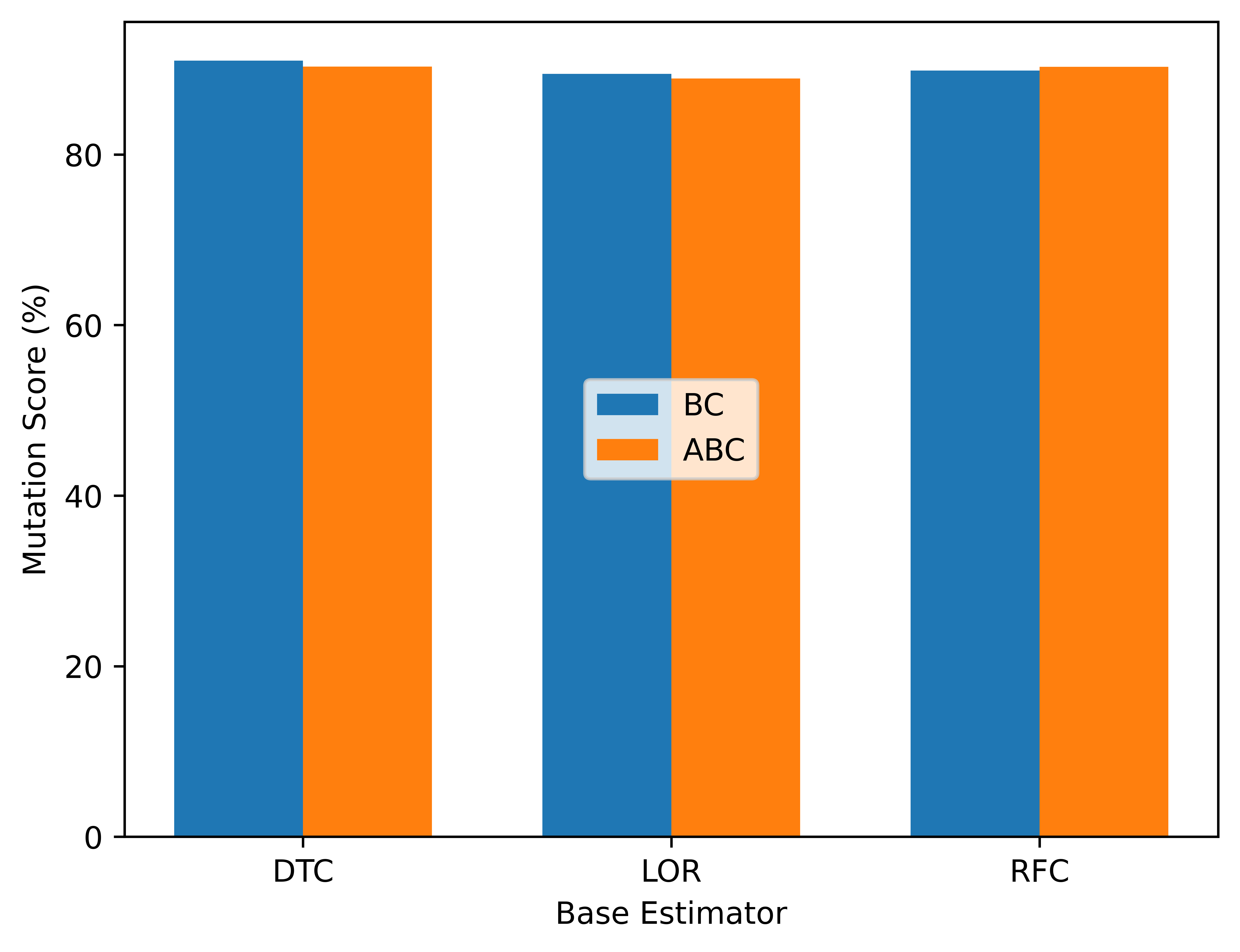}}

\subfloat[]{\includegraphics[width=0.33\textwidth]{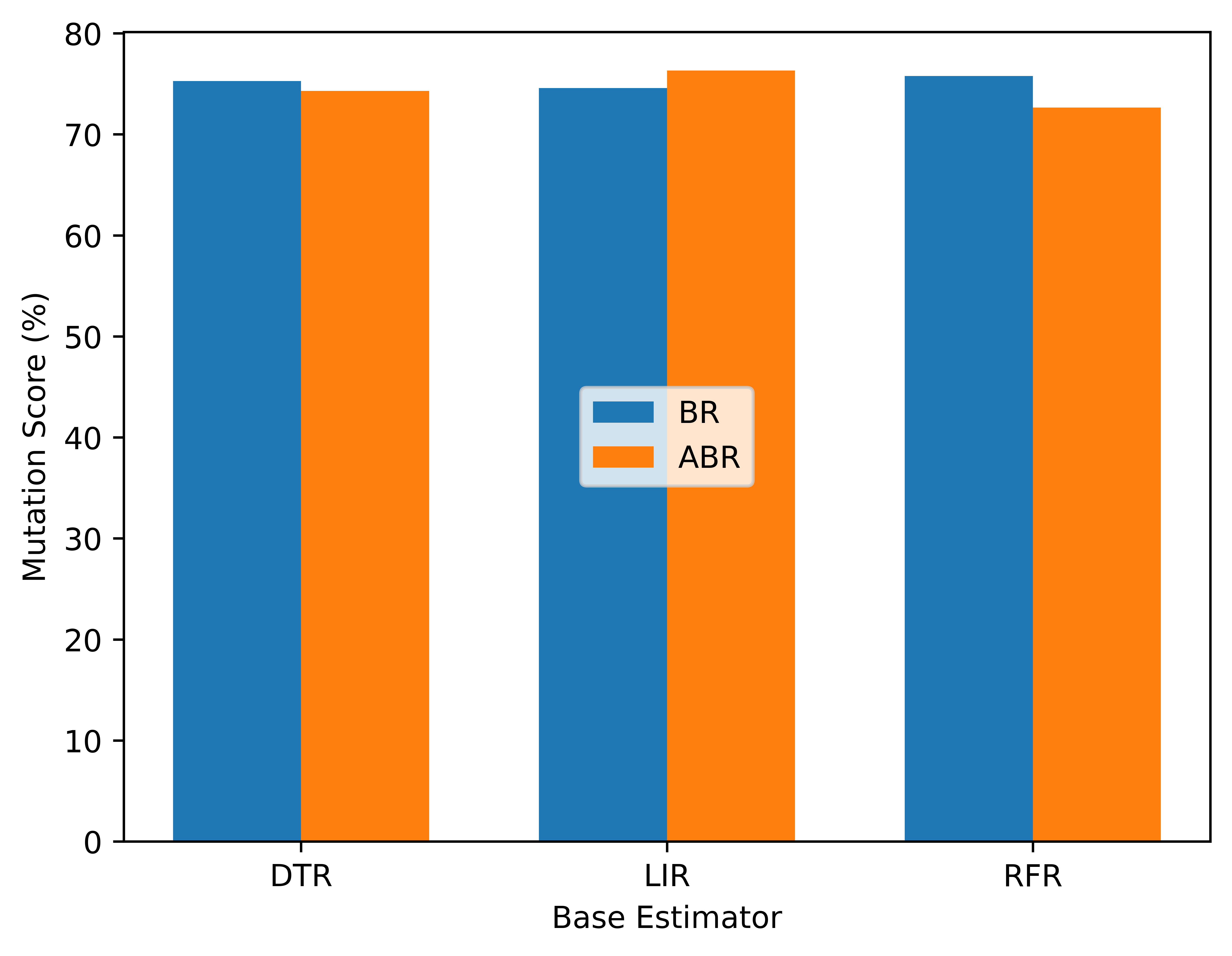}}
\subfloat[]{\includegraphics[width=0.33\textwidth]{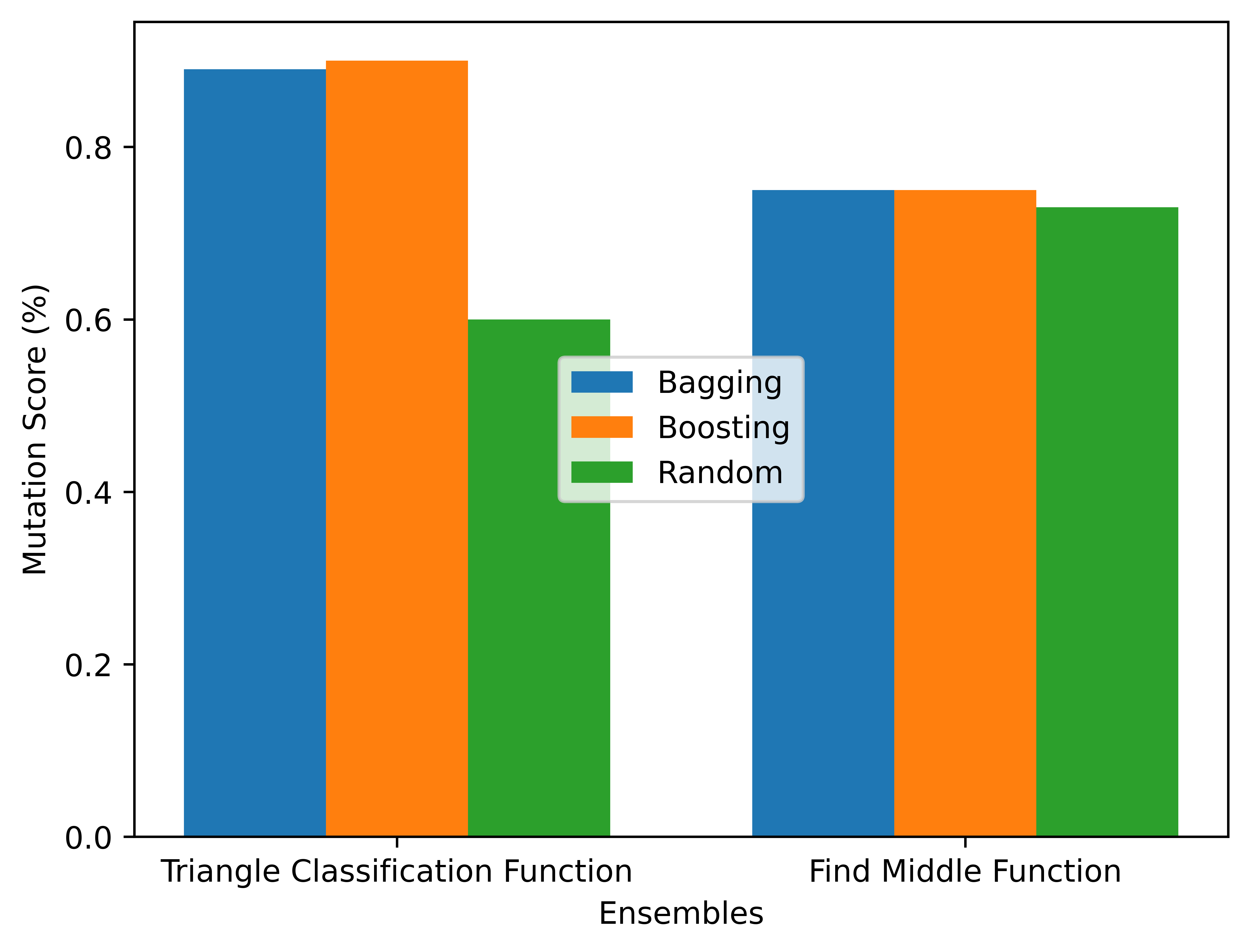}}
\subfloat[]{\includegraphics[width=0.33\textwidth]{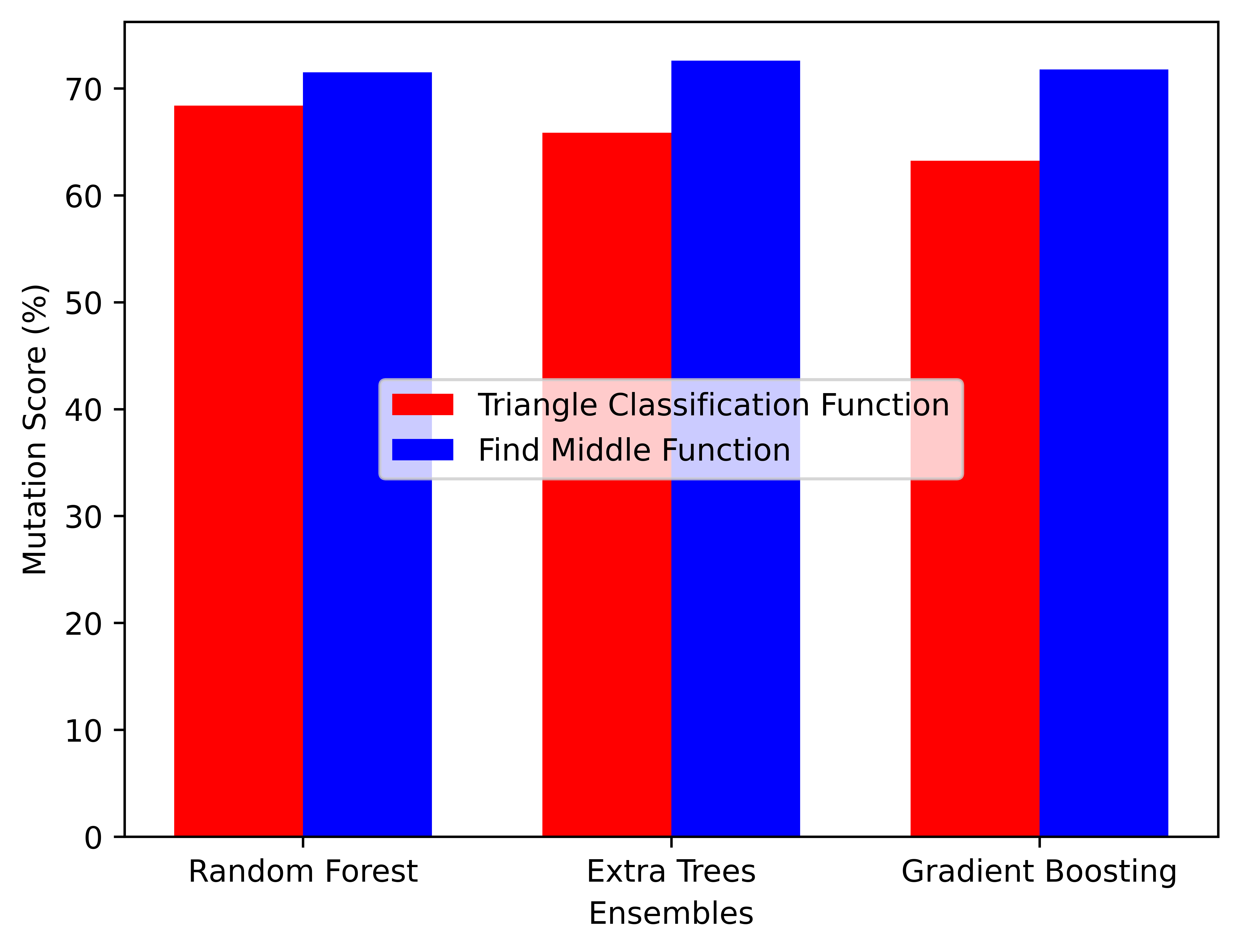}}
\caption{(a) Ensemble Performance - Triangle Classification, (b) Ensemble Performance - Find Middle, (c) Base Estimators Performance - Triangle Classification, (d) Base Estimators Performance - Find Middle, (e) Mutation Score Comparison - Bagging vs. Boosting, (f) Mutation Score Comparison of Random Forest, Extra Trees, and Gradient Boosting Methods.}
\label{fig:results}
\end{figure*}

\subsection{RQ1: Classification function} 
This research question aims to discover the most effective ensemble combinations for the triangle classification function. 

The triangle classification function undergoes mutation testing with $\mu$BERT, producing 14,693 mutants. To prevent inflated mutation scores, only 4,000 executable mutants are randomly selected. The effectiveness of various ensemble combinations in generating test suites is evaluated by comparing the number of killed mutants against those killed by a randomly generated test suite, as shown in Figure \ref{fig:results}(a).

The findings clearly indicate that test suites generated using the ensemble combinations killed a considerably higher number of mutants than the randomly generated test suite, highlighting their superior capability in detecting and handling mutants. Among the ensemble combinations, BC-DTC, ABC-DTC, ABC-RFC, and GBC has comparatively higher kill score than the other combinations. On the contrary, the ETC combination performs the least effectively among all the ensembles. However, even the worst-performing ensemble combination has killed a noticeably higher number of mutants compared to the randomly generated test suite. This result also indicates the potency of the proposed method for generating an effective test suite. Overall, ensemble combinations are performing at a similar level with little difference between their performance.

\subsection{RQ2: Numeric function}
The focus of this research question is to determine the most effective ensembles in generating test cases for the find middle function. 

The find middle function's $\mu$BERT generates 8,835 mutants, of which 3,736 are executable mutants selected for mutation testing. Figure~\ref{fig:results}(b) shows the total mutants killed by different ensembles for the find middle function, compared to the total mutants killed by a randomly generated test suite.

The results clearly show that the test suites generated by the BR-DTR, BR-RFR, ETR, ABR-LIR, and GBC combinations achieve the highest kill scores. These combinations demonstrate superior effectiveness in detecting and handling mutants, leading to higher overall mutation scores. Conversely, the combinations of RFR and ABR-RFR have comparatively lower mutant kill scores among these ensemble methods. 

Both Bagging and Boosting methods exhibit comparable mutation scores, with some fluctuations observed in the kill scores of various ensemble combinations. While all ensemble methods outperform randomly generated test suites in killing mutants, the differences in performance are less pronounced compared to the results obtained for the triangle classification function.

\subsection{RQ3: Better overall}


\textbf{\textit{Triangle Classification Function}.}
In the triangle classification function, three base estimators (DTC, RFC, and LOR) are utilized for two ensembles: BC and ABC. Figure~\ref{fig:results}(c) illustrates their performance in terms of mutation score for both BC and ABC. RFC exhibits slightly better performance with ABC, while DTC and LOR perform slightly better with BC compared to ABC. Overall, base estimators perform similarly in generating effective test suites for both BC and ABC, indicating the effectiveness of most ensemble combinations for the classification function. Additionally, all combinations outperform randomly generated test suites.

\textbf{\textit{Find Middle Function}.}
For the find middle function, we set DTR, LIR, and RFR as base estimators for both BR and ABR. Comparing base estimator performance for BR and ABR, its apparent that DTR and RFR outperform as base estimators for BR (Figure~\ref{fig:results}(d)), while LIR performs slightly better as a base estimator for ABR.


\textbf{\textit{Overall}.}
In Figures~\ref{fig:results}(c) and ~\ref{fig:results}2(d), a notable observation emerges where the performance of Decision Tree, Logistic Regression, and Random Forest as base estimators exhibits contrasting patterns in Bagging and Boosting methods for the two functions. For instance, Random Forest performs better as a base estimator for the Boosting method in the triangle classification function, whereas it performs better for the Bagging method in the find middle function. Similarly, Logistic Regression excels with the Bagging method in the triangle classification function, while Linear Regression performs better with the Boosting method for the find middle function.




Figure~\ref{fig:results}(f) shows that Random Forest, Extra Trees, and Gradient Boosting achieved higher mutation scores for the find middle function compared to the triangle classification function. Additionally, Figure~\ref{fig:results}(e) suggests that Boosting methods generally outperform other methods in terms of mutation score for both functions. While the disparity is more evident for the triangle classification function, where both Bagging and Boosting notably outperform random generation, for find middle, Boosting and Bagging still perform better, albeit less prominently.

\section{Discussion \& Conclusion}
\label{sec:Discussion}

Our results are consistent 
with several previous experiments that have indicated that Boosting outperforms Bagging for model inference~\cite{bauer1999empirical}. 
Boosting methods' superior performance could be attributed to their underlying approach. 
As mentioned earlier, LBT combines model inference and testing simultaneously. 
Boosting methods train models by assigning higher weights to instances that were incorrectly predicted in previous iterations. 
This incremental learning process aligns well with LBT's approach, where learning also occurs progressively.

In LBT, effective test cases are selected based on the disagreement among inferred models, where greater discrepancy indicates higher effectiveness. However, to avoid overfitting, inferred models must be cautious. Boosting tends to produce better ensembles than Bagging but is susceptible to noisy data, leading to overfitting by assigning excessive weights to noisy samples~\cite{maclin1997empirical}. Fortunately, our experiment lacked noisy samples, enhancing the performance of Boosting methods in generating effective test cases.

Our method performs better than random testing for the classification function but only slightly better for the numerical function. This difference could be due to the complexity of function specifications and the base estimators used for the model. For example, in a right scalene triangle (which is a subcategory of scalene), two sides need to be equal and the sum of the squares of those two sides needs to be equal to the square of the hypotenuse. Random generation may struggle with complex conditions like those of a right scalene triangle, but our approach, utilizing the Z3 solver to generate diverse test cases, shows potential for improved performance, especially with more complex programs. While our experiment focused on simple functions, future work will explore the potential of our approach by testing it on more complex systems.





\balance{}

\bibliographystyle{abbrv}

\bibliography{sigproc}  

\end{document}